\documentclass[aps,prd,groupedaddress,amsmath,amssymb,amsfonts,nofootinbib,preprint]{revtex4-1}

\usepackage{amssymb,amsfonts}
\usepackage{mathrsfs}
\usepackage{tikz}
\usepackage{lipsum}
\DeclareMathAlphabet{\mathpzc}{OT1}{pzc}{m}{it}
\usetikzlibrary{patterns}

\renewcommand{\d}{{\rm d}}

\begin{document}

\title{The Memory Effect for Particle Scattering in Even Spacetime Dimensions}

\author{David Garfinkle}
\email{garfinkl@oakland.edu}
\affiliation{Dept. of Physics, Oakland University, Rochester, MI 48309, USA}
\affiliation{Michigan Center for Theoretical Physics, Randall Laboratory of Physics, The University of Michigan, Ann Arbor, MI 48109-1120, USA}
\author{Stefan Hollands}
\email{stefan.hollands@uni-leipzig.de}
\affiliation{Institut fu¨r Theoretische Physik, Universit\" at Leipzig, Br\" uderstrasse 16, D-04103 Leipzig, Germany}
\author{Akihiro Ishibashi}
\email{akihiro@phys.kindai.ac.jp}
\affiliation{Department of Physics, Kindai University, Higashi-Osaka, 577-8502, Japan}
\author{Alexander Tolish}
\email{tolish@uchicago.edu}
\author{Robert M. Wald}
\email{rmwa@uchicago.edu}
\affiliation{Enrico Fermi Institute and Dept. of Physics, The University of Chicago, 5640 South Ellis Avenue, Chicago, IL 60637, USA}

\date{\today}

\begin{abstract}
We explicitly calculate the gravitational wave memory effect for classical point particle sources in linearized gravity off of an even dimensional Minkowski background. We show that there is no memory effect in $d>4$ dimensions, in agreement with the general analysis of \cite{H-I-W}.
\end{abstract}

\maketitle

\section{Introduction}

After a gravitational wave passes a wave detector composed of test particles at a large distance $r$ from the source, these test particles may exhibit a permanent change in their relative separation. This phenomenon is known as the \textit{gravitational wave memory effect} and was first recognized in the linear regime by Zel'dovich and Polnarev \cite{Zeldovich}. The memory effect remains an active topic of study, particularly with regard to the relationship between memory,``soft gravitons,'' and Bondi-Metzner-Sachs (BMS) supertranslations \cite{Strom1}-\cite{Kehagias}.

In a recent paper \cite{H-I-W}, it was shown that the memory effect vanishes to leading nontrivial order\footnote{The results of \cite{H-I-W} and the analysis of the present paper concern only the memory effect at the leading order in $1/r$ at which gravitational radiation can affect the motion of test particles. There are sub-dominant in $1/r$ effects that can lead to nontrivial displacements of test particles \cite{Chu1}-\cite{Chu2}, but we shall not consider such sub-dominant effects here.} in $1/r$ in all asymptotically flat spacetimes of even dimension $d>4$ that are stationary near spatial infinity and---after a burst of gravitational radiation---become (nearly) stationary again at late times at null infinity. The analysis of \cite{H-I-W} consisted of solving Einstein's equation near null infinity to find the metric form before and after the burst of radiation. However, it is not immediately evident from the general analysis of \cite{H-I-W} why there is a difference between dimension $d=4$ and dimensions $d>4$. The purpose of this paper is to calculate the memory effect for a simple example of point particle scattering in linearized gravity, where one can see more readily the differences between $d=4$ and $d>4$.

Although point particles are not compatible with the full nonlinear Einstein field equations \cite{Geroch}, they are allowed in the context of linearized perturbations\footnote{Point particles also can be treated beyond linear order in certain approximations, but regularization schemes are then required.} off of a fixed background spacetime, such as $d$-dimensional Minkowski spacetime. We wish to consider the following idealized classical scattering situation:
At early times, the particles are assumed to non-interacting and move with uniform velocity\footnote{Note that since the particles therefore come in from 
arbitrarily large distances at early times, 
the spacetime is not sufficiently stationary near spatial infinity for the analysis of \cite{H-I-W} to apply directly.}.
\vfill\eject
All of these ingoing particle worldlines meet at a single event $P$, where they can interact and/or be destroyed; new particles may also be created at $P$. However, conservation of the total stress-energy tensor requires that the total energy-momentum at $P$ is conserved. The outgoing particles are then assumed to be non-interacting and move with uniform velocity out to (timelike) infinity.

The retarded metric perturbation in the radiation zone can be found by convolving the stress-energy tensor of the source with the retarded Green's function of the linearized Einstein equation, expanding in powers of $1/r$ (where $r$ is the spatial distance from the scattering event to the observation point), and keeping only the dominant term. For the idealized scattering source we are considering, there is nontrivial radiation only at the retarded time, $U_P$, of the interaction event $P$. If the leading-order term with respect to $1/r$ of the curvature tensor possesses a derivative-of-a-delta-function term, $\delta'(U-U_P)$, then the geodesic deviation equation for test particles near null infinity gives rise to a permanent, finite change in particle separation---\textit{i.e.}, memory.

We have previously shown that there is such a memory effect in this context in $d=4$ spacetime dimensions \cite{T-W1}-\cite{T-B-G-W}. However, we shall show in this paper that in higher (even) dimensions, the leading order behavior of the Riemann tensor is of the form $\delta^{(d/2 -1)}(U-U_P)$. Integration of the geodesic deviation equation then yields that there is no permanent displacement of test particles. 

We will review the form of the retarded Green's functions for the wave equation and obtain the form of the retarded solution for scalar point particle scattering in Sec. \ref{Green}. We will then obtain the scalar and electromagnetic field analogs of the memory effect in Sec. \ref{Scalar}. We then calculate the gravitational field and memory in Sec. \ref{Gravitational}. Finally, in Sec. \ref{Multipole}, we will obtain formulas for memory in the limit of slow motion of the sources, where it can be readily understood why the gravitational memory effect vanishes for $d > 4$.

We work in geometrized units ($G=c=1$). Lower case Latin indices from the begining of the alphabet are abstract indices, and indices from the middle of the alphabet denote spatial components. Capital Latin indices are indices on the sphere. Indices $(i)$ and $(j)$ in parentheses are particle labels. $\delta^{(k)}$ denotes the $k^{\textnormal{th}}$ derivative of of a Dirac delta function, and $\delta_n$ is an $n$-dimensional delta function. Our conventions and notations for the signature, Riemann curvature tensor etc are the same as in \cite{Wald}.

\section{Retarded Solution to the Wave Equation with Particle Sources}\label{Green}

Consider the inhomogeneous scalar wave equation in $d$-dimensional Minkowski spacetime,
\begin{equation}
\partial^a\partial_a\varphi=-4\pi S \; .\label{wave}
\end{equation}
We will take $S$ to represent a system of scalar charged point particles that follow timelike inertial trajectories except at a single ``interaction vertex'' $P$, where they may interact and/or be created or destroyed. Let $(t,\mathbf{x})$ be a globally inertial coordinate system (GICS). Without loss of generality, we can choose our GICS so that $P$ is at the origin $(t=0,\mathbf{x}=0)$. Then $S$ takes the form
\begin{equation}
S(x)=
\sum_{(i)\textnormal{ in}}q_{(i)} \frac{\d\tau_{(i)}}{\d t} \delta_{d-1}\left(\mathbf{x}-\mathbf{y}_{(i)}(t)\right)\Theta(-t) + 
\sum_{(j)\textnormal{ out}}q_{(j)}\frac{\d\tau_{(j)}}{\d t} \delta_{d-1}\left(\mathbf{x}-\mathbf{y}_{(j)}(t)\right)\Theta(t) \; ,\label{scalarsource}
\end{equation}
where $q_{(i)}$ are the scalar charges of the particles as measured in their rest frame, 
and $(t,\mathbf{y}_{(i)} (t))$ are the particle worldlines  parametrized with the GICS time coordinate. 

We wish to find the retarded field $\varphi$ for such a source $S$. This will require convolving $S$ with the retarded Green's function $G$ for eq. \eqref{wave}:
\begin{equation}
\varphi(x)=4\pi\int \d^dx' G(x,x')S(x') \; .
\label{ret}
\end{equation}
For even\footnote{For odd dimensions, the Green's function in Minkowski spacetime takes a very different form and, in particular, does not have support on the light cone. Thus, the analysis of our paper applies only to even dimensions.} $d$, the retarded Green's function is given by
\begin{equation}
G(x,x')=\frac{1}{(2\pi)^{\frac{d}{2}-1}}\delta^{\left(\frac{d}{2}-2\right)}\left(\sigma^2(x,x')\right)
\Theta(t-t')
\end{equation}
(see, for example, Sec. 6.1 of \cite{Friedlander}), where 
\begin{equation}
\sigma^2(x,x')=-(t-t')^2+|\mathbf{x}-\mathbf{x'}|^2
\end{equation}
is the squared geodesic distance between points $x$ and $x'$. This can be rewritten as
\begin{equation}
G(x,x')=\frac{1}{2}\frac{1}{(2\pi)^{\frac{d}{2}-1}}\Theta(t-t')\left(-\frac{1}{\Xi}\frac{\partial}{\partial\Xi}\right)^{d/2-2}\left(\frac{\delta\left((t-t')-\Xi\right)}{\Xi}\right) \; ,
\label{Greenform}
\end{equation}
where, after taking the derivative, we substitute $\Xi=|\mathbf{x}-\mathbf{x'}|$. 

The composition of the retarded propagator with our distributional sources in eq.~\eqref{ret} is well defined, as one can see by a standard wave-front-set argument. Firstly, viewed as a bi-distribution, the retarded propagator is known to have on any globally hyperbolic spacetime $M$, in any dimension $d \ge 2$, the wave front set \cite{hormander}
${\rm WF}(G)$ consisting of those $(x,k_a^{},x', -k'_a)$ in the cotangent bundle $T^*(M \times M) \setminus 0$ minus the zero section 
for which there exists a future directed null-geodesic $\gamma$ connecting $x$ and $x'$ such that $k_a$ and $k'_a$ are co-tangent to- and parallel transported along $\gamma$. Secondly, for a single particle, the wave front set ${\rm WF}(S)$ consists of the co-normal bundle of the particle trajectory (i.e. the set of all non-zero covectors annihilating the tangent vector of the particle). Since the particle trajectory is timelike, it follows that there cannot exist a $(x, 0, x', -k'_a) \in {\rm WF}(G)$ and a $(x', p'_a) \in {\rm WF}(S)$ such that $p'_a-k'_a=0$. Consequently, by the wave front set calculus \cite{hormander}, the distributional product $G(x,x')
S(x')$ is well defined on any globally hyperbolic spacetime. Corresponding statements hold for electromagnetic sources $J_a$ and stress-energy sources $T_{ab}$.  Our calculations show furthermore that integration over all of $x'$ is admissible in Minkowski spacetime, i.e. there are no infra-red divergences. The case of more particles is treated in the same way, since the source is just the sum of contributions from the individual particles.

To obtain the retarded solution \eqref{ret}, we consider, first, a source $S_0$ corresponding to a single massive particle created at $P$ ``at rest,''
\textit{i.e.},
\begin{equation}
S_{\textnormal{out, } 0} (x)=q\delta_{d-1}(\mathbf{x})\Theta(t) \; .
\end{equation}
The retarded field of such a source is
\begin{align}
\varphi_{\textnormal{out, } 0} (x)&=2\pi q \frac{1}{(2\pi r)^{d/2-1}}\frac{\d^{d/2-2}}{\d U^{d/2-2}}\Theta(U)+O\left(\frac{1}{r^{d/2}}\right) \; , \label{varphi0}
\end{align}
where 
\begin{equation}
U=t-r
\end{equation}
is the retarded time coordinate. Thus, the leading order behavior of $\varphi_0$ is $(1/r)^{d/2 -1}$. 
We shall ignore the terms of sub-leading-order in $1/r$ throughout the remainder of this paper.

The field of a particle created with velocity $\mathbf{v}$ can be found by boosting eq. \eqref{varphi0}. For a particle following the worldline $(t,\mathbf{y}(t))$ with coordinate-velocity $\mathbf{v}=d\mathbf{y}/dt$, we obtain to leading order in $1/r$
\begin{equation}
\varphi_{\textnormal{out, }\mathbf{v}}(x)=2\pi q \frac{1}{(2\pi r)^{d/2-1}}
\frac{\d\tau}{\d t}
\frac{1}{1-\mathbf{\hat{r}\cdot v}}\frac{\d^{d/2-2}}{\d U^{d/2-2}}\Theta(U) \; ,\label{scalarfieldout}
\end{equation}
where $\mathbf{\hat{r}}=\mathbf{x}/r$ is the unit vector pointing from the scattering vertex to the observation point. These calculations can be repeated for a particle at rest that is ``destroyed'' at $P$
\begin{equation}
S_{\textnormal{in, }0}(x)=q \delta_{d-1}(\mathbf{x})\Theta(-t) \; ,
\end{equation}
in which case we find that at large distances the field of the boosted particle behaves like
\begin{equation}
\varphi_{\textnormal{in, }\mathbf{v}}(x)=2\pi q \frac{1}{(2\pi r)^{d/2-1}}
\frac{\d\tau}{\d t}
\frac{1}{1-\mathbf{\hat{r}\cdot v}}\frac{\d^{d/2-2}}{\d U^{d/2-2}}\Theta(-U) \; .\label{scalarfieldin}
\end{equation}

A general source of the form eq. \eqref{scalarsource} can be written as a linear superposition of such created and destroyed particles, so its field can be written as a superposition of individual fields like eqs. \eqref{scalarfieldout} and \eqref{scalarfieldin}.
Thus, we find that the retarded solution with source \eqref{scalarsource} is given by
\begin{equation}
\varphi_S(x) = \frac{2\pi}{(2\pi r)^{d/2-1}}\frac{\partial^{d/2-2}}{\partial U^{d/2-2}}\left(\Theta(U)\alpha(\mathbf{\hat{r}}) + \Theta(-U)\beta(\mathbf{\hat{r}})\right) \; .\label{scalarfield}
\end{equation}
to leading order in $1/r$, where
\begin{equation}
\alpha = \sum_{(i)\textnormal{, out}}
\frac{\d\tau_{(i)}}{\d t}
\frac{q_{(i)}}{1-\mathbf{\hat{r}\cdot v}_{(i)}}\, , \quad
\beta = \sum_{(j)\textnormal{, in}}
\frac{\d\tau_{(j)}}{\d t}
\frac{q_{(j)}}{1-\mathbf{\hat{r}\cdot v}_{(j)}}
\end{equation}

\section{Scalar and Electromagnetic Memory}\label{Scalar}

\subsection{Scalar Memory}

We now consider the effects of the scalar field \eqref{scalarfield} on test particles.
The scalar force on a test particle of mass $M_0$ and charge $Q$ is given by
\begin{equation}
f^a=Q\partial^a\varphi
\end{equation}
The leading order force at large distances for the field \eqref{scalarfield} is 
\begin{equation}
f^a(u,\mathbf{x})=-2\pi Q\frac{\alpha-\beta}{(2\pi r)^{d/2-1}}\frac{\d^{d/2-2}}{\d U^{d/2-2}}\delta(U)K^a \; ,\label{scalardiscon}
\end{equation}
where 
\begin{equation}
K^a= - \partial^a U \; .
\end{equation}
If the test particle is initially at rest, then its change in its momentum can be found by integrating \eqref{scalardiscon} with respect to time:
\begin{equation}
\Delta P^a(U)=\int_{-\infty}^{U}\d U' \, f^a(U',\mathbf{x}) = -2\pi Q\frac{\alpha-\beta}{(2\pi r)^{d/2-1}}\frac{\d^{d/2-2}}{\d U^{d/2-2}}\Theta(U)K^a \; .
\end{equation}

In $d=4$ dimensions, the change in momentum goes like $r^{-1}\Theta(U)K^a$, as was previously calculated for a particular case in \cite{T-W1}. Thus, in $d=4$ dimensions, a test particle will get a ``momentum kick'' as a result of the scalar radiation emitted by the interactions of the particles. Note that, generally, a test particle exposed to such a kick will experience a change in mass:\footnote{Changes in particle mass associated with relativistic scalar fields have been calculated before: see, for example, \cite{B-H-P} and \cite{H-P}.}
\begin{equation}
M_1^2=-\eta_{ab}(P_0^a+\Delta P^a)(P_0^b+\Delta P^b)=M_0^2-2P_0^a\Delta P_a =M_0^2-2Q(\alpha-\beta)\frac{M_0}{r}  \; ,
\end{equation}
up to terms of order $1/r^2$. 

In $d=6$ dimensions, the leading order momentum change of the test particle goes like $r^{-2}\delta(U)K^a$. Thus, to leading order, the test particle's momentum returns to its initial value after the scalar wave passes; there is no velocity kick or change of mass. For the idealized case considered here---where the radiation is instantaneous---there is no time for the test particle to move, so there is no change of position either. However, for a smoothed-out source---where the radiation acts over a finite time---the test particle would undergo a finite radial displacement.
In this sense, we obtain a scalar wave memory effect
for the position of a test particle when $d=6$.

In $d>6$ dimensions, both the momentum and position of the test particle return to their initial values after the passage of the wavefront. Even for a smoothed out source, the momentum change of the test particle averages to zero during the passage of the wave, so there is no permanent displacement. Thus, there is no scalar memory effect for $d>6$.

\subsection{Electromagnetic Memory}

A similar calculation can be done in classical electromagnetism. Maxwell's equations for the vector potential, $A^a$, in Lorenz gauge, $\partial_a A^a = 0$, take the form of the wave equation
\begin{equation}
\partial^a\partial_a A^b=-4\pi J^b \; ,\label{maxwell}
\end{equation}
where $J^a$ is the charge-current density, which satisfies the conservation law $\partial_a J^a=0$. We again consider ingoing and outgoing point charges that scatter at an event $P$, taken to be at the origin of our GICS. Each of the outgoing particles has a charge-current of the form
\begin{equation}
J_{(i)}^a=q_{(i)}\frac{\d \tau_{(i)}}{\d t}u_{(i)}^a\delta_{d-1}\left(\mathbf{x}-\mathbf{y}_{(i)}(t)\right)\Theta(t)
\end{equation}
whereas the incoming charges have charge-current
\begin{equation}
J_{(j)}^a=q_{(j)}\frac{\d\tau_{(j)}}{\d t}u_{(j)}^a\delta_{d-1}\left(\mathbf{x}-\mathbf{y}_{(j)}(t)\right)\Theta(-t)
\end{equation}
where $u_{(i,j)}^a$ are the normalized tangent vectors, $\tau_{(i,j)}$ are the proper times along the worldlines $(t,\mathbf{y}_{(i,j)})$
and $q_{(i,j)}$ are the electromagnetic charges as measured in the rest frame of the particle. 
Conservation of $J^a$ implies conservation of charge at the interaction vertex $P$, \textit{i.e.},
\begin{equation}
\sum_{(i)\textnormal{ out}} q_{(i)}=\sum_{(j)\textnormal{ in}} q_{(j)} \; .
\label{qcons}
\end{equation}

Using \eqref{scalarfield} on each GICS component of \eqref{maxwell}, we find that the retarded solution for $A_a$ is to leading order in $1/r$
\begin{align}
A^a(x)&=\frac{2\pi}{(2\pi r)^{d/2-1}}\frac{\partial^{d/2-2}}{\partial U^{d/2-2}}\left(\Theta(U)\alpha^a + \Theta(-U)\beta^a\right) \; \label{fourpotential}
\end{align}
where 
\begin{align}
\alpha^a(\mathbf{\hat{r}})&=\sum_{(i)\textnormal{ out}}
\frac{\d\tau_{(i)}}{\d t}
\frac{q_{(i)}u_{(i)}^a}{1-\mathbf{\hat{r}\cdot v}_{(i)}} \; ,\\
\beta^a(\mathbf{\hat{r}})&=\sum_{(j)\textnormal{ in}}
\frac{\d\tau_{(j)}}{\d t}
\frac{q_{(j)}u_{(j)}^a}{1-\mathbf{\hat{r}\cdot v}_{(j)}} \; .
\end{align}
The field tensor $F_{ab}=2\partial_{[a}A_{b]}$ is thus given to leading order in $1/r$ by
\begin{align}
F^{ab} =\frac{4\pi}{(2\pi r)^{d/2-1}}K^{[a}(\alpha^{b]}-\beta^{b]})\frac{\d^{d/2-2}}{\d U^{d/2-2}}\delta(U) \; .
\end{align}
Using conservation of charge \eqref{qcons}, it can be seen that 
\begin{equation}
K^{[a}(\alpha^{b]}-\beta^{b]})
=\sum_{{(i)}\textnormal{ in, out}}\frac{\d\tau_{(i)}}{\d t} \frac{\eta_{(i)}q_{(i)}}{1-\mathbf{\hat{r}\cdot v}_{(i)}}K^{[a} q^{b] c} u_{(i) c}  \; ,
\end{equation}
where $q_{ab}$ is the metric (projector) of the sphere
the factor $\eta_{(i)}$ equals $+1$ if particle $(i)$ is outgoing and $-1$ if it is ingoing. In particular, we have $F_{ab} K^b = 0$.

The force acting on a test particle with charge $Q$ and $4$-velocity $V^a$ is
\begin{equation}
f^a=QF^{ab}V_b \; .
\end{equation}
We assume that the test particle is initially at rest in our GICS, $V^a=t^a$. Then,
\begin{equation}
f^a(U,\mathbf{x})=\frac{2\pi Q}{(2\pi r)^{d/2-1}}\left[\sum_{(i)\textnormal{ in, out}}\frac{\eta_{(i)}q_{(i)}}{1-\mathbf{\hat{r}\cdot v}_{(i)}}
\frac{\d\tau_{(i)}}{\d t}
q^{ab}u_{(i) b}\right]\frac{\d^{d/2-2}}{\d U^{d/2-2}}\delta(U) \; .
\end{equation}
Its change in momentum is 
\begin{align}
\Delta P^a(U)&=\int_{-\infty}^U \d U' f^a(U',\mathbf{x})\\
&=\frac{2\pi Q}{(2\pi r)^{d/2-1}}\left[\sum_{(i)\textnormal{ in, out}}\frac{\eta_{(i)}q_{(i)}}{1-\mathbf{\hat{r}\cdot v}_{(i)}}
\frac{\d\tau_{(i)}}{\d t}
q^{ab}u_{(i)b}\right]\frac{\d^{d/2-2}}{\d U^{d/2-2}}\Theta(U) \; .
\end{align}
Since $f_a V^a = 0$, the electromagnetic force cannot produce a change in mass. Otherwise we obtain similar results to in the scalar case: For $d=4$, we find a velocity kick in a direction tangent to the sphere centered at $P$, as previously found in \cite{B-G1} and \cite{T-W1}. For $d=6$, for a smoothed-out source, we can obtain a finite displacement of the test particle's position tangent to the sphere. For $d > 6$, there is no electromagnetic memory effect.

\section{The Gravitational Field and Memory}\label{Gravitational}

We now turn our attention to gravitational memory arising from linearized gravitational perturbations $h_{ab}$ off of an even dimensional Minkowski background, with flat metric $\eta_{ab}$. In the harmonic gauge, $\partial^a \bar{h}_{ab} = 0$, the linearized Einstein equation takes the form
\begin{equation}
\partial^c\partial_c\bar{h}_{ab}=-16\pi T_{ab} \; ,\label{EFE}
\end{equation}
where $T_{ab}$ is the stress-energy tensor of the source and 
\begin{equation} 
\bar{h}_{ab}=h_{ab}-\frac{1}{2}\eta_{ab}h \; .
\end{equation}
We therefore find that  ${{\bar h}_{ab}}$ is given by
\begin{equation}
\bar h_{ab} (x) = 16 \pi  \int  \d^d x' \, (G \cdot I_{ab}{}^{c'd'})(x,x') T_{c' d'}(x')
\end{equation}
with $G$ denoting the retarded propagator of the scalar wave equation, and 
$I_{ab}{}^{c'd'}$ the bi-tensor of parallel transport, which is of course trivial in a GICS.
In fact, it follows from eqn. (\ref{Greenform}) that near null infinity, to lowest order in $1/r$ we have in a GICS $(x^\mu)$
\begin{equation}
{{\bar h}_{\mu \nu}}(r,U, \hat {\bf r}) = 8\pi \; {{(2\pi r)}^{-d/2+1}} \; 
\frac{\partial^{d/2-2}}{\partial U^{d/2-2}}\; \int \d^{d-1} {\bf y} \; {T_{\mu \nu}}(U+\hat{\bf r} \cdot {\bf y},{\bf y}) \; \; \; .
\label{hbar}
\end{equation}
Again, we consider ingoing and outgoing massive particles that interact only at a single event, $P$, taken to be at the origin of our GICS. 
The stress-energy of the $i$th outgoing particle of rest mass $m^{(i)}$ takes the form
\begin{equation}
T^{(i)}_{ab}=m^{(i)}u^{(i)}_au^{(i)}_b\delta_{d-1}(\mathbf{x}-\mathbf{y}^{(i)}(t))\frac{\d\tau^{(i)}}{\d t}\Theta(t) \; .\label{massSE}
\end{equation}
The stress-energy of a massive incoming particle $(j)$ takes the same form as eq. \eqref{massSE}, with $\Theta(t)$ replaced with $\Theta(-t)$. The full stress-energy tensor is
\begin{equation}
T_{ab}=\sum_{(i)\textnormal{ out}} T_{ab}^{(i)}+\sum_{(j)\textnormal{ in}} T_{ab}^{(j)}
\end{equation}
and conservation of stress-energy, $\partial^a T_{ab}=0,$ implies
\begin{equation}
\sum_{(i)\textnormal{ out}} m^{(i)}u_a^{(i)} = \sum_{(j)\textnormal{ in}} m^{(j)}u_a^{(j)} \; .
\end{equation}
Applying eq. \eqref{scalarfield} to each GICS component of \eqref{EFE}, we obtain to leading order in $1/r$
\begin{equation}
h_{ab}=\frac{8\pi}{(2\pi r)^{d/2-1}}\frac{\partial^{d/2-2}}{\partial U^{d/2-2}}\left(\Theta(U)\alpha_{ab} + \Theta(-U)\beta_{ab}\right) \; ,
\label{reth}
\end{equation}
where
\begin{equation}
\alpha_{ab}(\mathbf{\hat{r}})=\sum_{(i)\textnormal{ out}}\frac{m^{(i)}}{1-\mathbf{\hat{r}\cdot v}_{(i)}}\frac{\d\tau^{(i)}}{\d t}\left(u_a^{(i)}u_b^{(i)}+\frac{1}{d-2}\eta_{ab}\right) 
\end{equation}
and
\begin{equation}
\beta_{ab}(\mathbf{\hat{r}})=\sum_{(j)\textnormal{ in}}\frac{m^{(j)}}{1-\mathbf{\hat{r}\cdot v}_{(j)}}\frac{\d\tau^{(j)}}{\d t}\left(u_a^{(j)}u_b^{(j)}+\frac{1}{d-2}\eta_{ab}\right) \; .
\end{equation}
The linearized Riemann tensor around $\eta_{ab}$ is in any number of dimensions and any gauge 
\begin{equation}
R_{acbd} = \partial_{a} \partial_{[d} h_{b]c} - \partial_{c} \partial_{[d} h_{b]a} .
\label{Riem1}
\end{equation}
The linearized Riemann tensor computed from \eqref{reth} is
\begin{equation}
R_{abcd}=\frac{8\pi}{(2\pi r)^{d/2-1}}K_{[a}\Delta_{b][c}K_{d]}\frac{\d^{d/2-1}}{\d U^{d/2-1}}\delta(U) \; ,\label{Riemann}
\end{equation}
where
\begin{equation}
\Delta_{ab}=2\sum_{(i)\textnormal{ in, out}}\frac{\eta_{(i)}m_{(i)}}{1-\mathbf{\hat{r}\cdot v}_{(i)}}\frac{\d\tau^{(i)}}{\d t} \left\{ 
q_{ac}u^c_{(i)}q_{bd}u^d_{(i)} + \frac{1}{d-2}q_{ab}\right\} \; .\label{S-tensor}
\end{equation}
Again, $\eta_{(i)}$ is $+1$ for outgoing and $-1$ for incoming particles. 
The non-vanishing components of this expression for $\Delta_a{}^b$  in the 
coordinates $(U,r,z^A)$ (where $z^A$ are coordinates on $S^{d-2}$) can be rewritten as
\begin{equation}
\label{Deltadef1}
\Delta_{A}{}^B=\left(D_A D^B -\frac{1}{d-2}\delta_A{}^B D^C D_C\right)T \; .
\end{equation}
where $D_A$ is the covariant derivative on the unit round sphere, where the indices in this equation are raised 
with the inverse metric of the {\em unit} round sphere, and
\begin{equation}
T(\mathbf{\hat{r}}) = 2\sum_{(i)\textnormal{ in, out}}\eta_{(i)}\left(E_{(i)}-\mathbf{\hat{r}\cdot p}_{(i)}\right)\ln\left(E_{(i)}-\mathbf{\hat{r}\cdot p}_{(i)}\right) \; .
\label{supertrans}
\end{equation}
Here
\begin{equation}
(E,\mathbf{p})=m\frac{\d t}{\d \tau}\left(1,\mathbf{v}\right)\label{d-velocity}
\end{equation}
is a particle's relativistic momentum (\textit{i.e.}, its ``$d$-momentum'').

The relative motion of test particles (and thus memory) is described by the geodesic deviation equation. If two test particles are initially at rest in the GICS $(t,x^i)$ -- so that their $4$-velocities are both $t^a$ -- and spatially separated by the displacement vector $\xi^i$, then their relative motion will be governed by
\begin{equation}
\frac{\d^2\xi_i}{\d t^2} = -R_{i0k0}\xi^k \; .\label{gde2}
\end{equation}
Going to a GICS $(t,x^i)$ in eq.~\eqref{Riem1}, applying the relation $\partial _i {\bar h}_{\mu \nu} = - {\hat r}_i\partial _t {\bar h}_{\mu \nu}$ valid to lowest 
order in $1/r$ in a GICS [seen e.g. from \eqref{hbar}], and using the harmonic gauge condition gives after some straightforward algebra~\footnote{
Alternatively, use \eqref{Riemann} in a GICS.} 
\begin{equation}
{R_{i0k0}} = -{\frac 1 2} \frac{\partial^2}{\partial t^2} {\cal P}[{\bar h}]_{ik}
\label{Riem2}
\end{equation}
where the projection operator $\cal P$ is defined for any symmetric tensor $A_{ik}$ by
\begin{equation}
{{\cal P}[A]}_{ik} \equiv {{q_i}^m}{{q_k}^n}{A_{mn}} - \;  {\frac 1 {(d-2)}} {q^{mn}}{A_{mn}} {q_{ik}} \; \; \; .
\label{Pdef}
\end{equation}
(Here ${{q_i}^k}$, the projector to the sphere, is given in Cartesian coordinates by ${{q_i}^k} \equiv {{\delta _i}^k} - {{\hat r}_i}{{\hat r}^k}$).
That is, $\cal P$ projects a symmetric tensor to be orthogonal to $\hat {\bf r}$ and trace-free.  This is the analog for all even dimensions of the usual notion in 4 dimensions that gravitational radiation is described by the transverse, traceless part of the metric.  Now integrating eqn. (\ref{gde2}) twice with respect to time and using eqn. (\ref{Riem2}) we find that the change in displacement of the two test particles is in a GICS,
\begin{equation}
\Delta {\xi^i} = {\frac 1 2} {{\cal P}[\Delta \bar h]}_{k}{}^i \xi ^k . 
\label{memgen}
\end{equation}
Upon inserting \eqref{reth} we find that the change in $\xi^a$ for test particles in the presence of a scattering process is
\begin{equation}
\Delta \xi^i(U)=\frac{2\pi}{(2\pi r)^{d/2-1}}\Delta_k^{\;\; i}\frac{\d^{d/2-2}}{\d U^{d/2-2}}\Theta(U)\xi^k \; ,
\end{equation}
where $\Delta_{ik}$ are the spatial Cartesian coordinates of our displacement tensor \eqref{S-tensor}, given alternatively by \eqref{Deltadef1}
in terms of the coordinates $(U,r,z^A)$.

For $d=4$, we have
\begin{equation}
\Delta \xi^i=\frac{1}{r}\Delta^{\;\; i}_{ k}\xi^k \; .
\end{equation}
Thus, there is a nontrivial memory effect in $4$ spacetime dimensions. As discussed in \cite{H-I-W}, the quantity $T$ appearing in \eqref{supertrans} describes the supertranslation associated with the memory effect.

However, for $d>4$, the change in separation goes like $\delta(U)$ or derivatives of $\delta(U)$. The test particles return to their original relative displacement, and there is no memory effect.

\section{Memory in the Slow Motion Limit}\label{Multipole}

Further insight into the absence of gravitational memory for $d > 4$ can be seen from consideration of radiation in the slow-motion limit of the source. To analyze this, there is no need to restrict consideration to particle sources, and we shall not make this restriction below except where stated. To leading order in the velocity of the source, we may neglect the variation of the retarded time over the source. For the scalar field \eqref{wave} with source $S$, to lowest order in source velocity and leading order in $1/r$, we thereby obtain
\begin{eqnarray}
\varphi(U,\mathbf{x}) &=& \frac{2\pi}{(2\pi r)^{d/2-1}}\frac{\d^{d/2-2}}{\d U^{d/2-2}}\int \d^{d-1}{\bf x}' \, S(U,\mathbf{x'})  \nonumber \\
&=& \frac{2\pi}{(2\pi r)^{d/2-1}}\frac{\d^{d/2-2}\Omega}{\d U^{d/2-2}} \; ,
\end{eqnarray}
where $\Omega = \Omega(U)$ denotes the monopole moment of the source at retarded time $U$.
Thus, the leading order contribution to scalar radiation comes from variation of the monopole moment. For electromagnetic radiation, we similarly obtain for the spatial components, $A_i$, of the vector potential,
\begin{equation}
A_i (U,\mathbf{x}) = \frac{2\pi}{(2\pi r)^{d/2-1}}\frac{\d^{d/2-2}}{\d U^{d/2-2}}\int \d^{d-1} {\bf x}' \, J_i(U,\mathbf{x'}) \; .
\end{equation}
However, using conservation of $J^a$, we have
\begin{equation}
\int \d^{d-1}{\bf x}' \, J^i(U,\mathbf{x'}) = \int \d^{d-1}{\bf x}' \, J^j(U,\mathbf{x'}) \partial_j x'^i = \frac{\d}{\d U}\int \d^{d-1} {\bf x}' J^0(U,\mathbf{x'})x'^i = \frac{\d p^i}{\d U} 
\end{equation}
where $p^i = p^i (U)$ is the electric dipole moment of the source.
Thus, we obtain
\begin{equation}
A_i (U) = \frac{2\pi}{(2\pi r)^{d/2-1}}\frac{\d^{d/2-1}p_i}{\d U^{d/2-1}} \; ,
\end{equation}
and the dominant form of electromagnetic radiation in the slow motion limit is electric dipole radiation. 

Similarly, in the gravitational case, we have
\begin{equation}
\int \d^{d-1}{\bf x}' \, T_{ij}(U,\mathbf{x'}) = \frac{1}{2}\frac{\d^2}{\d U^2}\int \d^{d-1}{\bf x}' \, T_{00}(U,\mathbf{x'})x'_ix'_j \equiv \frac{1}{2}\frac{\d^2I_{ij}}{\d U^2} \; ,
\end{equation}
where $I_{ij}=I_{ij}(U)$ is the inertia tensor. The dominant contribution to the spatial components of the metric perturbation is thus
\begin{equation}
\bar{h}_{ij} (U) = \frac{4\pi}{(2\pi r)^{d/2-1}}\frac{\d^{d/2} I_{ij}}{\d U^{d/2}} \; \; .
\end{equation}
Note however that gravitational radiation depends only on the projected tensor ${\cal P} [\bar{h}]_{ij}$.
The inertia tensor is decomposed as
\begin{equation}
{I_{ij}} = {Q_{ij}} + {\frac 1 {d-1}} {{I^k}_k}{\delta _{ij}} \; \; \; ,
\end{equation}  
where $Q_{ij}$ is trace-free and is called the quadrupole tensor.  
Since $ {\cal P}[{\delta}]_{ij}=0$, it follows that gravitational radiation depends only on the quadrupole tensor.  Thus
the dominant form of gravitational radiation in the slow motion limit is (polar parity) quadrupole radiation.

The gravitational memory effect is determined by the change in $\bar{h}_{ij}$ between asymptotically early and late times. In $d=4$ spacetime dimensions, this will be given by the change in $\d^2 I_{ij}/\d U^2$. If the matter is moving in from infinity in an inertial manner at early times and moving out to infinity in an inertial manner at late times, it is not difficult to produce a change in this quantity. In particular, for particle sources, we have 
\begin{equation}
I_{ij}(t)=\sum_{(i)} m^{(i)} x^{(i)}_i x^{(i)}_j \; ,
\end{equation}
and if the particles are in inertial motion, we have
\begin{equation}
\frac{\d^2{I}_{ij}}{\d U^2}= 2 \sum_{(i)} m^{(i)} v^{(i)}_iv^{(i)}_j  \; .
\end{equation}
In general, this quantity is nonvanishing, and its value for incoming particles need not equal its value for outgoing particles. 

By contrast, for $d>4$, the memory effect in the slow motion limit is given by the change in derivatives of $I_{ij}$ of higher order than second. However, for matter in inertial motion, we have
\begin{equation}
\frac{\d^3 I_{ij}}{\d U^3}= 0 \; .
\end{equation}
Thus, one can thereby see that there can be no memory effect in the slow motion limit when $d > 4$.

\bigskip

\noindent
{\bf Acknowledgements}

The work of D.G. was supported by NSF grant PHY 15-05565 to Oakland University. The research of A.I. was supported by Japan Society for the Promotion of Science (JSPS)  KAKENHI Grants
No. 15K05092 and No. 26400280. The research of A.T. and R.M.W. was supported by NSF grant PHY 15-05124 to the University
of Chicago.

\end{document}